\documentclass[%
reprint,
amsmath,amssymb,
aps,
prl
]{revtex4-2}
\usepackage[normalem]{ulem}

\usepackage{graphicx}
\usepackage{bm}
\usepackage{xcolor}

\begin{document}	
	\preprint{APS/123-QED}
	
	\title{Dynamic Fingerprint of Controlled Structural Disorder in Artificial Spin Lattices} 
	
	\author{Vinayak Shantaram Bhat}\email[]{Authors to whom correspondence should be addressed: vbhat@udel.edu, mbj@udel.edu}
	\affiliation{Department of Physics and Astronomy, University of Delaware, Newark, DE 19716, USA}%
	\author{M Benjamin Jungfleisch}\email[]{Authors to whom correspondence should be addressed: vbhat@udel.edu, mbj@udel.edu}
	\affiliation{Department of Physics and Astronomy, University of Delaware, Newark, DE 19716, USA}%
	\date{\today}
	\begin{abstract}
		Investigating the emergence of complexity in disordered interacting systems, central to fields like spin glass physics, remains challenging due to difficulties in systematic experimental tuning. We introduce a tunable artificial spin lattice platform to directly probe the connection between controlled structural disorder and collective spin-wave dynamics. By precisely varying positional and rotational randomness in Ni\(_{81}\)Fe\(_{19}\) nanobar arrays from periodic to  random, we map the evolution from discrete spectral modes to a complex, dense manifold. Crucially, we establish a quantitative correlation between information-theoretic measures of static disorder and the dynamic spectral complexity  derived from the GHz spin-wave response. This correlation provides a dynamic fingerprint of an increasingly complex energy landscape resulting from tuned disorder. Furthermore, thermal probe via thermal Brillouin light scattering  reveal significantly richer microstates diversity in disordered states than driven probe using broadband ferromagnetic resonance. Our work presents a unique experimental testbed for studying how the ingredients of glassy physics manifest in high-frequency dynamics, offering quantitative insights into the onset of complexity in interacting nanomagnet systems.
	\end{abstract}
	
	\maketitle
	
	\maketitle
	Understanding emergent phenomena in disordered interacting systems, like the notoriously complex spin glass state \cite{BinderYoungRMP1986,ParisiPhysLettA1979}, is a cornerstone of condensed matter physics. Canonical spin glasses offer rich physics \cite{mydosh1993spin,nordblad1998experiments}, but their inherent chemical disorder limits systematic tuning of parameters to explore connections between microscopic randomness and macroscopic behavior \cite{EdwardsAndersonJPC1975}. Artificial spin lattices (ASLs), including artificial spin ice \cite{wang2006artificial, NisoliMoessnerSchifferRMP2013,SkjaervoNatRevPhys2020}, provide platforms with geometric control \cite{HeydermanStampsJPD2013}, yet quantitatively linking tunable quenched disorder [obtained by predetermined variations in the structural arrangement of the Ni\(_{81}\)Fe\(_{19}\) (permalloy) nanobars] to resulting collective dynamics, particularly for glassy features \cite{saccone2019towards,FarhanNatPhys2013}, remains challenging. Establishing such links is crucial for fundamental understanding and for applications leveraging complexity, such as neuromorphic reservoir computing \cite{gartside2022reconfigurable}.	 Here, we address this by utilizing ASLs with precisely engineered and independently tunable positional and rotational disorder as an experimental testbed. This unique control, often intractable in conventional disordered materials, allows systematic investigation of how glassy physics ingredients manifest in high-frequency dynamic response. \\
	By fabricating permalloy nanobar arrays from perfect periodicity to full randomness, we map the collective GHz spin-wave spectrum's evolution using vector network analyzer based broadband ferromagnetic resonance (VNA-FMR), thermal Brillouin light scattering (BLS) spectroscopies \cite{demokritov2001brillouin}, and micromagnetic simulations \cite{donahue2002oommf}.
	  Our goals are: (i) disentangling positional versus rotational disorder effects on spin-wave dynamics, (ii) establishing quantitative relationships between static structural disorder metrics ($S_{\text{config}}$, $S_{\text{connect}}$) and emergent dynamic spectral complexity ($S_{\text{spectral}}$), and (iii) interpreting spectral evolution as a dynamic fingerprint of the underlying energy landscape complexity. We show positional disorder primarily lifts spectral degeneracies via modified dipolar coupling, while rotational disorder is the dominant driver for proliferating modes into a dense spectral manifold. We establish quantitative correlations linking static entropy measures to spectral complexity, revealing how dynamic richness is encoded in structural randomness. Furthermore, FMR and BLS comparison highlights thermal probes' ability to access greater microstate diversity in disordered states. This work provides insights into how different randomness forms sculpt wave propagation and offers a controlled platform for studying complexity onset relevant to glassy physics and engineering tailored dynamics.

	\begin{figure}[b]
		\includegraphics [width=0.49\textwidth]{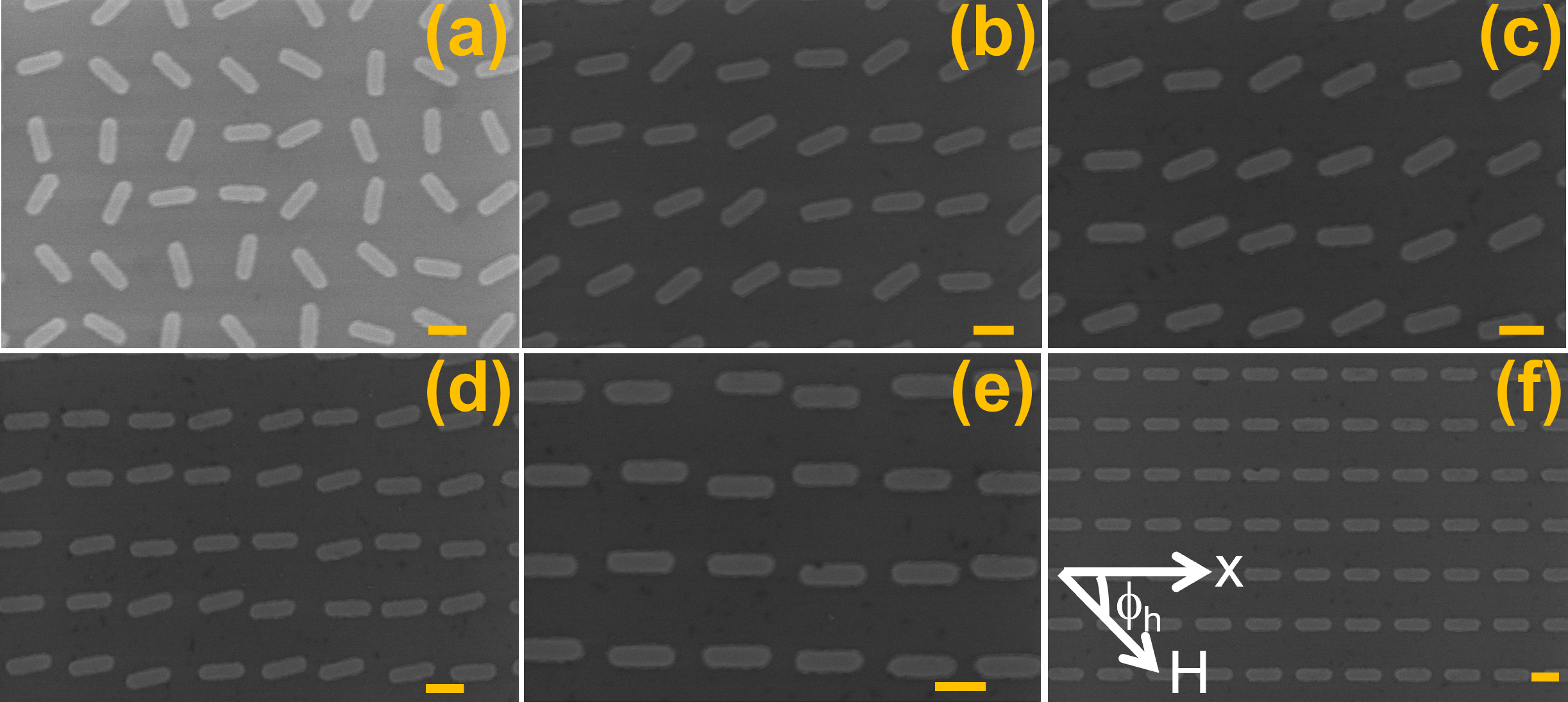}
		\centering
		\caption{Scanning electron microscopy images  of Permalloy ASL on Si for samples (a) \textit{S-A}, (b) \textit{S-B}, (c) \textit{S-C}, (d) \textit{S-D}, (e) \textit{S-E} and (f) \textit{S-F}, with orange scale bars representing 200 nm.}\label{Fig1}
	\end{figure}
    
\begin{figure}[t]
		\includegraphics [width=0.49\textwidth]{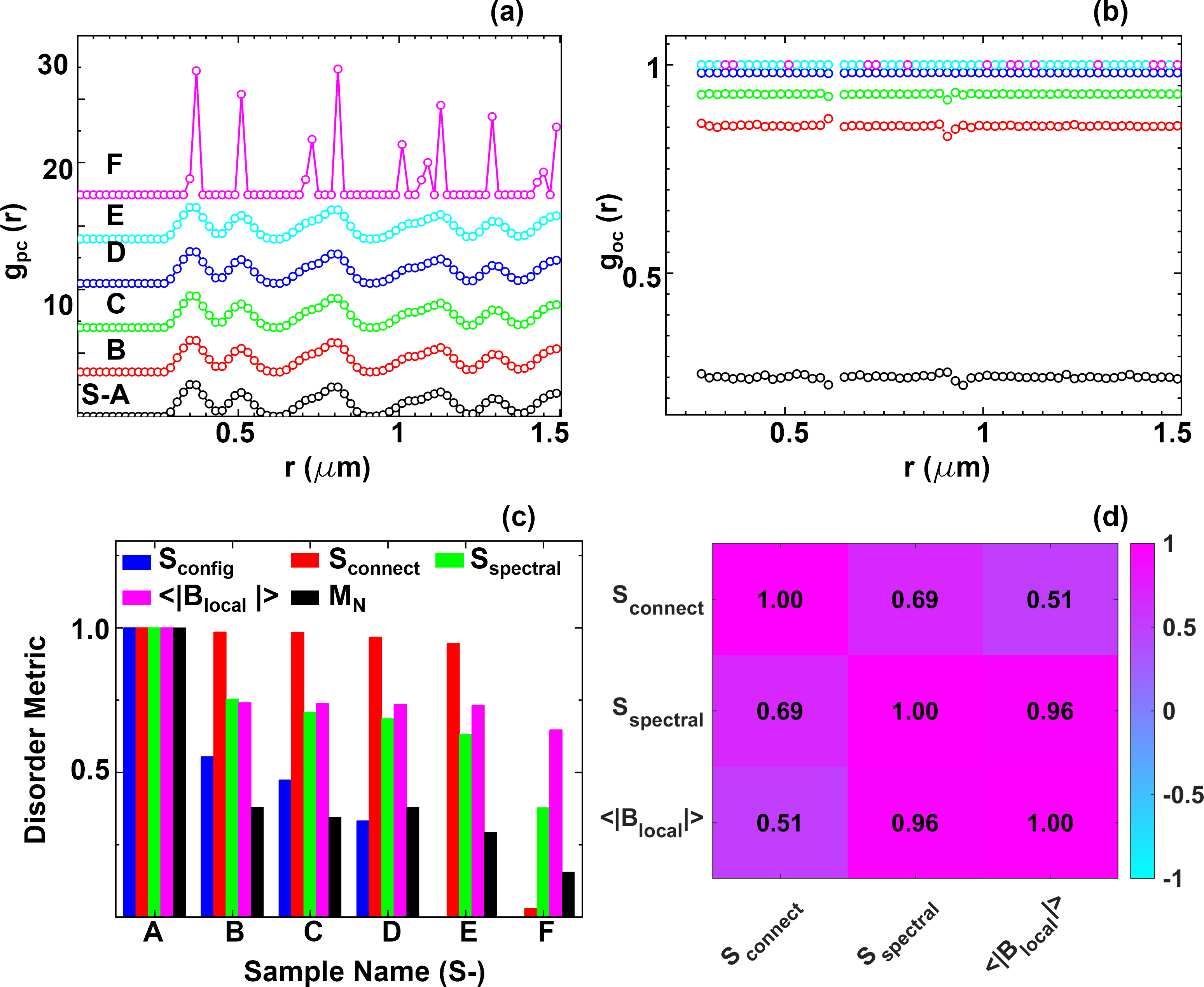} 
		\centering
		\caption{ (a) Pair correlation $g_{pc}(r)$ with curves vertically offset (samples \textit{S-F} top, \textit{S-A} bottom). (b) Orientational correlation $g_{oc}(r)$. Black, red, green, blue, cyan, and magenta symbols in (a) and (b) correspond to samples \textit{S-A}, \textit{S-B}, \textit{S-C}, \textit{S-D}, \textit{S-E}, and \textit{S-F}, respectively. (c) Comparison of metrics, each normalized with respect to the value observed in sample \textit{S-A}: configurational entropy ($S_{\text{config}}$, blue), connectivity entropy ($S_{\text{connect}}$, red), spectral entropy ($S_{\text{spectral}}$, green) for $H = -2$ kOe and $\phi_h = 1^\circ$, average local dipolar field ($\langle |B_{\text{local}}| \rangle$, magenta), and  key spectral feature count ($M_{N}$, black)  at $H = -2$ kOe and $\phi_h = 1^\circ$ in micromagnetic simulations. (d) Pearson correlation heatmap for $S_{\text{connect}}$, $S_{\text{spectral}}$, $\langle |B_{\text{local}}| \rangle$ across samples \textit{S-A} to \textit{S-F}, showing strength  of correlation.}\label{Fig2}
	\end{figure}

	To explore controlled disorder in ASLs, we fabricated samples with controlled structural variations [Fig.~\ref{Fig1}]. The baseline ASL (\textit{S-F}) consisted of permalloy nanobars (260\,nm \(\times\) 80\,nm \(\times\) 20\,nm), aligned along the \(x\)-axis with a 360\,nm nominal lattice constant. Positional disorder was introduced by varying nanobar separations via a Gaussian distribution (with mean 0 nm and standard deviation of 100 nm), ensuring no nanobar overlap. In-plane rotational randomness used a discrete uniform distribution from 0 to \(\phi\), producing six sample variants [Fig.~\ref{Fig1}] each with over 60,000 nanobars: \(\phi = 0^\circ\) (S-E), \(15^\circ\) (S-D), \(30^\circ\) (\textit{S-C}), \(45^\circ\) (\textit{S-B}), and \(360^\circ\) (\textit{S-A}, fully random), plus the periodic \textit{S-F}. Two versions of each sample (\textit{S-A} to \textit{S-F}) were prepared: one on a coplanar waveguide signal line for VNA-FMR, another on Si substrate for BLS.\\
To quantitatively characterize these structural variations and the resulting interaction landscape, we first examined spatial [Fig. \ref{Fig2}(a)] and orientational [Fig. \ref{Fig2}(b)] correlations across the sample series. Spatially resolved information was obtained from the pair correlation function $g_{pc}(r)$ and the orientational correlation function $g_{oc}(r)$. The pair correlation function $g_{pc}(r)$ measures the probability of finding a nanobar center at a distance $r$ from a reference nanobar center, relative to that for a random distribution at the same number density  $\rho = N/A$,  where \( N \) is the total number of nanobars and \( A \) is the total area they occupy. The function $g_{pc}(r)$ is computed as the ratio of the observed number of pairs to the number of pairs expected for a random distribution within the same area \cite{HansenMcDonald, McQuarrieStatMech}:
\begin{equation}
	g_{pc}(r) = \frac{N_{\text{obs}}(r)}{N_{\text{ideal}}(r)}, \quad \text{where } N_{\text{ideal}}(r) = \frac{N \rho}{2} A_{\text{shell}}(r)
	\label{eq:gpc_main_text_revised}
\end{equation}
Here, $N_{\text{obs}}(r)$ is the observed number of nanobar pairs within an annular shell $A_{\text{shell}}(r)$ corresponding to distance $r$, and $A_{\text{shell}}(r)$ is is the exact area from $r$ to $r+dr$ [see Supplemental Material (SM)  calculation details for disorder and complexity metrics)].

	Orientational alignment was probed via the orientational correlation function $g_{oc}(r)$, based on the second Legendre polynomial $P_2(x) = (3x^2 - 1)/2$ \cite{de1993physics, chaikin1995principles, bray1984theory,HansenMcDonaldTheorySimpleLiquids}:
	\begin{equation}
		g_{oc}(r) = \left\langle \frac{3\cos^2(\delta\theta_{ij}) - 1}{2} \right\rangle_r,
		\label{eq:goc_main_text}
	\end{equation}
	with $\delta\theta_{ij}$ the acute angle between orientation axes of nanobars $i$ and $j$, and averaging over all pairs at separation $r$ (see SM for details). The periodic array (\textit{S-F}) shows sharp peaks in $g_{pc}(r)$ and $g_{oc}(r) \approx 1$, indicating long-range positional and orientational order. Positional disorder (\textit{S-E}) broadens $g_{pc}(r)$ into damped oscillations but leaves $g_{oc}(r) \approx 1$. Introducing rotational disorder (\textit{S-D} to \textit{S-A}) maintains short-range positional order but reduces $g_{oc}(r)$, with $g_{oc}(r) \approx 0.25$ in \textit{S-A}, reflecting orientational decoherence that scales with disorder angle $\phi$.
	
	To further quantify disorder, we computed four global metrics. First, the configurational entropy $S_{\text{config}}$ captures orientational randomness. We binned nanobar angles $\theta_i \in [0, 2\pi)$ into $M=100$ intervals, constructed the distribution $p_k$, and computed Shannon entropy \cite{shannon1948} $S_{\text{config}} = -\sum p_k \ln p_k$, which increases from zero (\textit{S-F}/\textit{S-E}) to a maximum in \textit{S-A} [Fig.~\ref{Fig2}(c)].
	
	Second, the average dipolar field $\langle |B_{\text{local}}| \rangle$ estimates interaction strength. Each nanobar $j$ (moment $\vec{m}_j$ aligned along $\theta_j$ at $r_j$) contributes a dipolar field at nanobar $i$ via $\vec{B}_{ij} \propto (3(\vec{m}_j \cdot \hat{r}_{ij})\hat{r}_{ij} - \vec{m}_j)/r_{ij}^3$, summed over neighbors to yield $\vec{B}_{\text{local}, i}$ \cite{coey2010magnetism}. Averaging $|\vec{B}_{\text{local}, i}|$ over all $i$ gives $\langle |B_{\text{local}}| \rangle$, which grows from \textit{S-F} to \textit{S-A}, consistent with tighter local packing in disordered arrays.
	
	Third, to distinguish \textit{S-E} from \textit{S-F}, we define a connectivity entropy $S_{\text{connect}}$ based on weighted local interactions. For each nanobar $i$, the weighted degree $d_{w,i}$ sums alignment- and distance-dependent weights $w_{ij} \propto 1/r_{ij}^3$ over nearby $j$ within an effective radius $r_{eff}(i, j)$ , with weights modulated by relative orientation. The entropy $S_{\text{connect}} = -\sum p(d_w) \ln p(d_w)$, computed from the distribution $p(d_w)$, captures increasing environmental heterogeneity with positional and rotational disorder (see SM).
	
\begin{figure}[t]
		\includegraphics[width=0.49\textwidth]{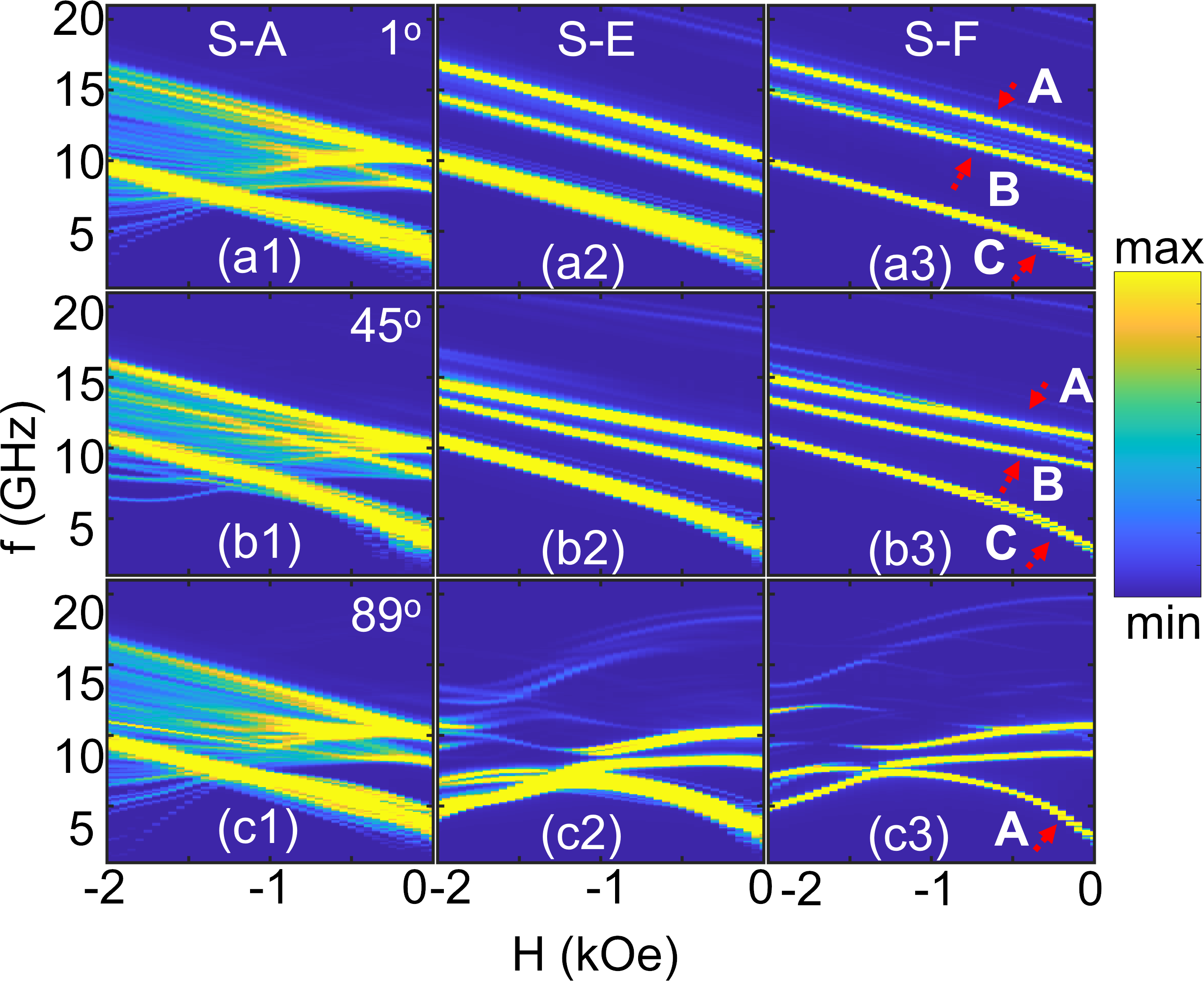}
		\caption{Integrated micromagnetic simulation power spectra with color scale indicating square of spin precession amplitudes (blue min, yellow max). Sub-figures follow the \textquotedblleft LN\textquotedblright naming format, where \textquotedblleft L\textquotedblright represents letters and \textquotedblleft N\textquotedblright represents numbers. Sub-figures with the same letter correspond to the same magnetic field angle (indicated on the right of the first column), while those with the same number correspond to the same sample (indicated at the top of the first row). Arrows/labels in panels (a3), (b3), and (c3) indicate the spin-wave branch designations for sample \textit{S-F} at $\phi_{h} = 1^\circ$, $45^\circ$, and $89^\circ$, respectively.}\label{Fig3}
	\end{figure}

	Fourth, the spectral entropy $S_{\text{spectral}}$ quantifies dynamic complexity from the simulated micromagnetic power spectra [Fig.~\ref{Fig3}]. Normalized power $p(f)$ over frequency bins yields $S_{\text{spectral}} = -\sum p(f) \ln p(f)$ (see SM for details). Higher values indicate broader or more fragmented spectral features. 
    
    Together, these metrics quantify the progression from ordered to disordered behavior across \textit{S-F} to \textit{S-A}, capturing trends in orientational disorder ($S_{\text{config}}$), interaction strength ($\langle |B_{\text{local}}| \rangle$), structural heterogeneity ($S_{\text{connect}}$), and dynamical complexity ($S_{\text{spectral}}$).
    
	Figure~\ref{Fig2}(c) quantitatively tracks tuned structural parameters across samples \textit{S-F} to \textit{S-A}, alongside dynamic complexity from spectral entropy ($S_{\text{spectral}}$). Metrics reveal a striking, coordinated evolution: as the engineered disorder increases from periodic \textit{S-F} to random \textit{S-A}, there is a systematic increase in configurational entropy ($S_{\text{config}}$), connectivity entropy ($S_{\text{connect}}$), average local dipolar field strength ($\langle |B_{\text{local}}| \rangle$), and the simulated spectral entropy ($S_{\text{spectral}}$). Sample \textit{S-F} represents the ordered baseline with minimal entropy values ($S_{\text{config}}=0$), while S-A exhibits the maximum achieved disorder across these metrics. Intermediate samples (\textit{S-E} to \textit{S-B}) map the progression from controlled positional and rotational randomness. \\
	Quantitative analysis [Fig.~\ref{Fig2}(d)] reveals correlations \cite{SnedecorCochranStatMethods,KutnerAppliedLinearModels} between these metrics [see SM for details]. $S_{\text{connect}}$ shows a moderate positive correlation with the $\langle |B_{\text{local}}| \rangle$ (Pearson coefficient $\approx 0.51$). However, strong positive correlations exist between $S_{\text{connect}}$ and $S_{\text{spectral}}$ (coefficient $\approx 0.69$) and also between $\langle |B_{\text{local}}| \rangle$ and $S_{\text{spectral}}$ (coefficient $\approx 0.96$). This indicates that both structural heterogeneity ($S_{\text{connect}}$) and average interaction strength ($\langle |B_{\text{local}}| \rangle$) serve as strong indicators of the dynamic complexity. A multiple linear regression analysis confirms that $S_{\text{connect}}$ and $\langle |B_{\text{local}}| \rangle$ together account for a significant portion of the variance in $S_{\text{spectral}}$ (adjusted $R^2 \approx 0.948$, F-statistic = 46.92, p= 0.0054). \\
	The controlled design of our sample series enables a clear disentanglement of the effects of positional and rotational disorder on the dynamic response. Introducing positional disorder alone (\textit{S-F} $\rightarrow$ \textit{S-E}) enhances structural heterogeneity ($S_{\text{connect}}$) and mean interaction strength ($\langle |B_{\text{local}}| \rangle$), yielding a modest rise in spectral complexity ($S_{\text{spectral}}$) [Fig.~\ref{Fig2}(c)]. In contrast, adding and increasing rotational disorder (\textit{S-D} $\rightarrow$ \textit{S-A}), captured by configurational entropy ($S_{\text{config}}$), leads to a substantially larger increase in $S_{\text{spectral}}$ [SM Fig.~S6]. This demonstrates that orientational randomness is a key driver in proliferating dynamic modes and enhancing spectral complexity beyond what positional disorder alone can produce. These results directly parallel concepts from spin glass physics, where isolating the roles of random bonds and random anisotropies—emulated here by positional and rotational disorder, respectively—is essential to understanding emergent complexity in frustrated systems \cite{BinderYoungRMP1986}. Our platform thus provides a tunable framework to systematically explore how distinct forms of quenched disorder shape dynamics in systems with rugged energy landscapes.

	The consequences of this quantified structural variation are directly evident in the simulated spectra [representative data in Fig.~\ref{Fig3}]. We observe an evolution from discrete, well-defined spin-wave modes in the periodic sample \textit{S-F} to increasingly complex and significantly broadened spectra as disorder increases towards sample \textit{S-A} [see SM Fig. S3 for simulated spin-wave spectra for all the samples]. Specifically, for sample \textit{S-F} at \(H = -2\,\text{kOe}\), three prominent modes—\textit{A}, \textit{B}, and \textit{C}—are observed. Spin-precession amplitude maps reveal that mode \textit{A} originates from nanobars aligned with the magnetic field, while modes \textit{B} and \textit{C} are lower harmonics, with power concentrated along two edges. Mode \textit{C} exhibits edge-mode characteristics, localized predominantly at the semicircular edges (SM Fig. S4) \cite{NeusserAdvMater2009}. As the magnetic field angle increases, these branches shift to lower frequencies; at \(89^\circ\), mode \textit{A} reaches 4.9\,GHz. With increasing rotational disorder from \textit{S-F} to \textit{S-A}, the number of spin-wave branches rises, spanning 4.9 to 17.1\,GHz (Fig.~\ref{Fig3}). \\
    Sample \textit{S-E}, despite lacking rotational disorder, exhibits lifted degeneracy because the modes are now  smeared out over a frequency range and are not as degenerate as in sample \textit{S-F}
	 [Fig.~\ref{Fig3}(a2)-(a3)]. This suggests that positional disorder alone influences dipolar interactions \cite{ostman2018interaction}. The observed increased number of spectral key features [black bars in Fig. \ref{Fig2}(c), see SM for details] between sample \textit{S-E} and \textit{S-A} correlates strongly with the increasing configurational entropy $S_{\text{config}}$ [Fig. \ref{Fig2}(c)], indicating that  orientational disorder leads to a wider range of resonant responses. Moreover, the increased spectral complexity and the smearing of distinct modes align with the increasing interaction heterogeneity quantified by $S_{\text{connect}}$ and the larger average interaction strength $\langle |B_{\text{local}}| \rangle$ found in the more disordered samples [Fig. \ref{Fig2}(c)]. The broadening is consistent with enhanced inhomogeneous broadening arising from the more heterogeneous distribution of local dipolar fields in the disordered samples \cite{KalarickalJAP2006}.
	
\begin{figure}[htp]
		\includegraphics [width=0.49\textwidth]{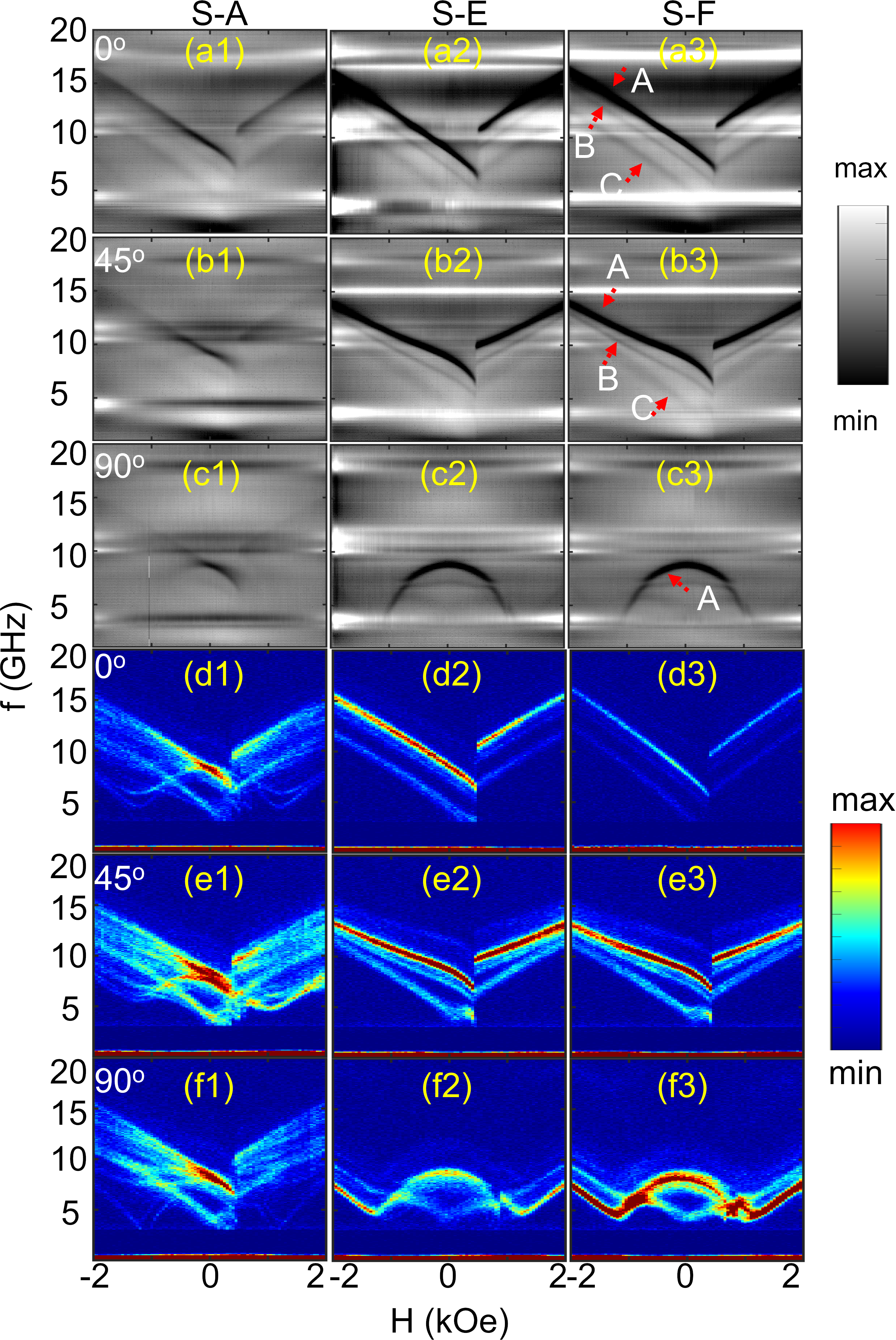}
		\caption{Grayscale VNA-FMR spin-wave spectra and jet color-scale thermal BLS spectra, showing high-contrast lines that represent spin-wave branches for samples at three different in-plane field angles. Sub-figures follow the \textquotedblleft LN\textquotedblright naming format, where \textquotedblleft L\textquotedblright represents letters and \textquotedblleft N\textquotedblright represents numbers. Sub-figures with the same letter correspond to the same magnetic field angle (indicated on the left of the first column), while those with the same number correspond to the same sample (indicated at the top of the first row). The arrows and labels in panels (a3), (b3), and (c3) indicate the spin-wave branch designations for sample \textit{S-F} at $\phi = 0^\circ$, $45^\circ$, and $90^\circ$, respectively.}\label{Fig4}
	\end{figure}

	To corroborate simulated findings and above mentioned disorder metrics, we performed VNA-FMR [Fig. \ref{Fig4}(a1)-(c3)], which revealed a systematic evolution of spin-wave branches as functions of magnetic field and angle. As the field angle increases, the spin-wave branches shift to lower frequencies as expected due to lower magnetic field component along the nanobar long axis, which also matches with mode analysis done using Smit-Beljer formulation [SM Fig. S1] \cite{smit1955philips, suhl1955ferromagnetic, lendinez2021emergent, vittoria2023magnetics}. Moreover, the linewidth of the main FMR mode is increased from 0.69 GHz for sample \textit{S-F} to 0.75 GHz \textit{S-E} for \(\phi_h = 45^\circ\) at $H = -2$~kOe. For an ideal uniform distribution of randomness, the FMR spectra of sample \textit{S-A} should exhibit an isotropic distribution of spin-wave branches across all field angles [compare Fig.~\ref{Fig3}(c1) and Fig.~\ref{Fig4}(c1)]. However, the observed spectra deviate from this expectation.
	
	To gain further insight into the spin-wave spectrum, particularly the thermally populated modes potentially sensitive to the increasing configurational entropy ($S_{\text{config}}$), we employed micro-focused BLS. Figures~\ref{Fig4}(d1)-(f3) present BLS spectra for representative samples across different applied field angles ($\phi_h = 0^\circ, 45^\circ, 90^\circ$). In the ordered sample \textit{S-F}, distinct modes are observed, including a prominent high-frequency branch. As disorder increases towards sample \textit{S-A} (corresponding to increasing $S_{\text{config}}$ and connectivity entropy $S_{\text{connect}}$), a clear proliferation of modes occurs, particularly at lower frequencies. This emergence of additional branches is attributed to the growing diversity of nanobar orientations relative to the applied field, each supporting different resonance conditions. Consistent with simulations and FMR, BLS also reveals that the primary mode (A) frequency is slightly lower in the positionally disordered sample \textit{S-E} compared to the periodic \textit{S-F}, indicating that positional disorder alone modifies the effective dipolar fields. Notably, in the fully disordered sample \textit{S-A}, the dense spectrum appears qualitatively similar across different field angles, suggesting a trend towards dynamic isotropy, a feature also observed in simulations. This contrasts with the ordered sample \textit{S-F}, which exhibits strong anisotropy. Furthermore, thermally excited BLS reveals a significantly higher number of detectable spin-wave modes compared to FMR. Notably, the dominant \textit{W}-shaped resonance branch at \(\phi_h = 90^\circ\) is clearly resolved via BLS across the samples, unlike in FMR, where mode intensities depend strongly on the magnetization’s alignment with the microwave field. This disparity stems from fundamental differences in excitation mechanisms: BLS spectra [Fig.~\ref{Fig4}(d1)-(f3)] reflect thermally activated spin waves, excited independently of sample orientation relative to a microwave field, ensuring relatively uniform detection efficiency across all modes. In contrast, FMR relies on inductive coupling, which weakens for modes misaligned with the microwave torque.
	
	In conclusion, utilizing artificial spin lattices with precisely engineered disorder as a tunable experimental testbed, we investigated the emergence of dynamic complexity relevant to glassy physics. Systematically varying positional and rotational randomness allowed us to map the transformation of the collective GHz spin-wave spectrum from discrete modes to a dense, complex manifold. This spectral evolution acts as a \textit{dynamic fingerprint} of an increasingly complex energy landscape shaped by the tuned disorder. Our central finding is the establishment of a quantitative correlation between static structural entropy measures ($S_{\text{config}}$, $S_{\text{connect}}$) and the dynamic spectral complexity ($S_{\text{spectral}}$), demonstrating that dynamic richness is directly encoded in quantifiable static randomness. This work provides a powerful platform for studying complexity onset and offers quantitative insights essential for designing a tailored magnonic and neuromorphic computing architecture \cite{ChumakNatPhys2015}.
	
	\section{Supplemental Material} 
	The Supplemental Material includes the VNA-FMR characterization results of the patterned and plain Py film at various in-plane angles, detailed micromagnetic simulation methodology, results on the simulated local power maps at $H = -2$~kOe, and discussion on correlation analysis, disorder entropy metrics, and multi-regression analysis method utilized.

	\section{Acknowledgment}
	This material is based upon work supported by the National Science Foundation under Grant No. 2339475. The authors acknowledge the use of facilities and instrumentation supported by NSF through the University of Delaware Materials Research Science and Engineering Center, DMR-2011824. The supercomputing time was provided by DARWIN (Delaware Advanced Research Workforce and Innovation Network), which is supported by NSF Grant No. MRI-1919839. MBJ acknowledges the JSPS Invitational Fellowship for Researcher in Japan.
	\section{Author's Contribution}
	VB and MBJ conceived the experiment. VB designed and fabricated the samples, performed the VNA and BLS measurements, micromagnetic simulations and data analysis. VB and MBJ wrote the manuscript.   \\

	\section{References}
	\bibliographystyle{apsrev4-2} 
	\bibliography{Bibliography1}
\end{document}


	\title{Supplemental Material: Dynamic Fingerprint of Controlled Structural Disorder in Artificial Spin Lattices}
	
	\author{Vinayak Shantaram Bhat}
	\email{vbhat@udel.edu}
	\affiliation{Department of Physics and Astronomy, University of Delaware, Newark, DE 19716, USA}%
	\author{M Benjamin Jungfleisch}
	\email{mbj@udel.edu}
	\affiliation{Department of Physics and Astronomy, University of Delaware, Newark, DE 19716, USA}%
	
	\date{\today}
	\maketitle
	
	This Supplemental Material includes the VNA-FMR characterization results of the patterned and plain  Py films at various in-plane field angles. Furthermore, we provide a detailed discussion of the micromagnetic simulation methodology, followed by results on the simulated local power maps at $H = -2$~kOe for sample \textit{S-F}. We also provide details on various disorder entropy metrics, correlation functions, correlation analysis, and regression analysis.
	
	\subsection{Smit-Beljers  Formalism} 
	
	To analytically understand the contributions of the Py layers in the observed spin-wave (SW) spectra of the studied artificial spin lattice (ASL) structures, we employed the Smit-Beljers-Suhl formalism. This formalism provides the conditions for ferromagnetic resonance in a single nanobar by describing the relationship between the angular resonance frequency and the free energy of the sample in spherical polar coordinates \cite{smit1955philips, suhl1955ferromagnetic, lendinez2021emergent, vittoria2023magnetics}:
	
	\begin{equation}\label{eq:sm_SBF-0_v4} 
		\left( \frac{\omega}{\gamma}  \right)^{2}  = \frac{1}{M_{eff}^{2} \sin^{2}\theta_{m}} \left(  \frac{\partial^2 F}{\partial \theta_{m}^{2}}\frac{\partial^2 F}{\partial \phi^{2}}  -  \left(\frac{\partial^{2} F}{\partial \theta_{m} \partial \phi}\right)^2  \right). 
	\end{equation}
	Here, $\theta_{h}$ ($\theta_{m}$) and $\phi_{h}$ ($\phi_{m}$), $\frac{\gamma}{2\pi}$ ( = 28 GHz/T), and $\mu_{o}$  represent polar and azimuth angles for applied field (effective magnetization), gyromagnetic ratio, and permeability of free space, respectively. The free energy expression for the nanobar, considering only demagnetization and Zeeman energy contributions, is:
	
	\begin{multline}\label{eq:sm_SBF-1_v4} 
		F = -\mu_{o}M_{eff}H\left[ \cos\theta_{h} \cos\theta_{m} + \sin\theta_{h} \sin\theta_{m} \cos(\phi_{h}-\phi_{m})                \right] +\\ \dfrac{\mu_{o}}{2}M_{eff}^{2}(N_{x}\sin^{2}\theta_{m}\cos^{2}\phi_{m}+ N_{y}\sin^{2}\phi_{m}\sin^{2}\theta_{m}+N_{z}\cos^{2}\theta_{m}). 
	\end{multline}
	For an in-plane magnetic field (XY-plane), $\theta_{m} = \theta_{h} = \frac{\pi}{2}$. The value of $\phi_{m}$ was determined by minimizing the free energy, which is done by setting $\frac{\partial F}{\partial \phi_{m}} = 0$. Using the Smit-Beljers formalism, the resonance frequency for a single nanobar can then be expressed as:
	\begin{widetext}
		\begin{equation}\label{eq:sm_SBF-2_v4} 
			f = \left(\frac{\gamma}{2\pi}\right)\mu_{o}\sqrt{\left[H\cos\left(\phi_{h} - \phi_{m} \right) + M_{eff}(N_{y} - N_{x})\cos(2\phi_{m})  \right] \times \left[H\cos\left(\phi_{h} - \phi_{m} \right) + M_{eff}(N_{z} - N_{x}\cos^{2}(\phi_{m}) - N_{y}\sin^{2}(\phi_{m}))  \right]}.  
		\end{equation}
	\end{widetext}
	
	Here, $N_{x} = 0.0614$, $N_{y} = 0.209$, and $N_{z} = 0.729$ are the demagnetization factors along the $x$-, $y$-, and $z$-direction determined using Ref. \cite{aharoni1998demagnetizing} for a 260 nm (length) $\times$ 80 nm (width) $\times$ 20 (thickness) nm nanobar.  
	
	The yellow dashed lines in Fig.~\ref{SI1} depict the resonance frequencies calculated using the Smit-Beljers formalism for 20\,nm Py layers as a function of applied field, assuming \(M_\text{eff} = 10\)\,kOe (equivalent to $M_s = 8 \times 10^5$ A/m), across various in-plane angles (Fig.~\ref{SI1}). The model accurately reproduces the experimental data for sample \textit{S-F}: at \(\phi_h = 0^\circ\) and \(\phi_h = 45^\circ\), the yellow line aligns closely with the spin-wave branches in \textit{S-F} and \textit{S-E}. However, from \textit{S-E} to \textit{S-A}, the branches shift to lower (higher) frequencies relative to the yellow line for \(\phi_h = 0^\circ\) (\(\phi_h = 45^\circ\)), reflecting increased rotational disorder. This disorder reduces the number of nanobars aligned with the applied field as randomness increases from \textit{S-E} to \textit{S-A}. The Smit-Beljers model also accounts for the \textit{W}-shaped profiles of branch \textit{A} at \(\phi_h = 90^\circ\), driven by field-dependent magnetization along the nanobar width. At high fields, magnetization aligns with the applied field, yielding a positive slope \(\frac{df}{dH}\). At lower fields, \(\frac{df}{dH}\) depends on \(\phi_m\), derived from free-energy minimization. However, branch intensities decrease significantly from \textit{S-D} to \textit{S-A} due to fewer nanobars aligning with the CPW signal line, rendering microstate-specific ferromagnetic resonance less identifiable.

	\subsection{Micromagnetic Simulation Details}
	To gain a more detailed understanding of the dynamics,  we took a subset of square ASL arrays for each sample occupying a nominal  area of  3.5 microns $\times$ 3.5 microns  on a 5 nm ($x$-axis) $\times$ 5 nm ($y$-axis) $\times$ 20 nm grid (that is, only one layer along the $z$-direction) and simulated  using the Object Oriented Micromagnetic Framework (OOMMF) \cite{donahue2002oommf} [Fig. 3 in main text].  The Py parameters used in the simulations  were as follows: exchange constant $A = 1.3 \times 10^{-11} \,\mathrm{J\,m^{-1}}$, saturation magnetization  $M_{\mathrm{S}}= 8 \times 10^{5} \,\mathrm{A\,m^{-1}}$ (= 10 kOe), magnetocrystalline anisotropy constant $K = 0$, gyromagnetic ratio $\gamma = 2.211 \times 10^{5} \,\mathrm{m\,A^{-1}\,s^{-1}}$, and dimensionless damping coefficient $\alpha = 0.01$. \\
	First, we obtained DC magnetization maps for each applied field value by sweeping the field from $-12$ kOe to +12 kOe  at a fixed in-plane angle $\phi$.
	The equilibrium magnetization configurations, at a given field and angle, from the above mentioned  DC hysteresis simulations were then used as inputs for magnetodynamic simulations. A dynamical simulation was conducted by considering a Gaussian magnetic field pulse of 20 mT  amplitude and 2.5 ps full width at half maximum along the $z$-direction (perpendicular to the film  plane). We recorded 512 values of magnetization vectors for each grid pixel at 20 ps time steps. The perpendicular component of magnetization was then logged as a function of $x, y$ and the time step. Subsequently, a fast Fourier transformation (FFT) was performed on the magnetization of each grid point along the time axis to obtain the resonance spectrum. The absorbed power for each grid value was obtained by squaring the absolute value of FFT amplitude, that is, $[\text{Re}(\text{FFT}(m_z))]^2$. The analysis provided a map of spin-precessional amplitudes at a selected frequency.  The power spectra  were obtained by integrating over all grid points for each frequency step. We show spatial distributions ($x,y$ maps) of square of spin-precessional amplitudes in Fig. \ref{SI2}.
		\renewcommand{\thefigure}{S\arabic{figure}}
	\begin{figure}[htp]
		\includegraphics [width=0.99\textwidth]{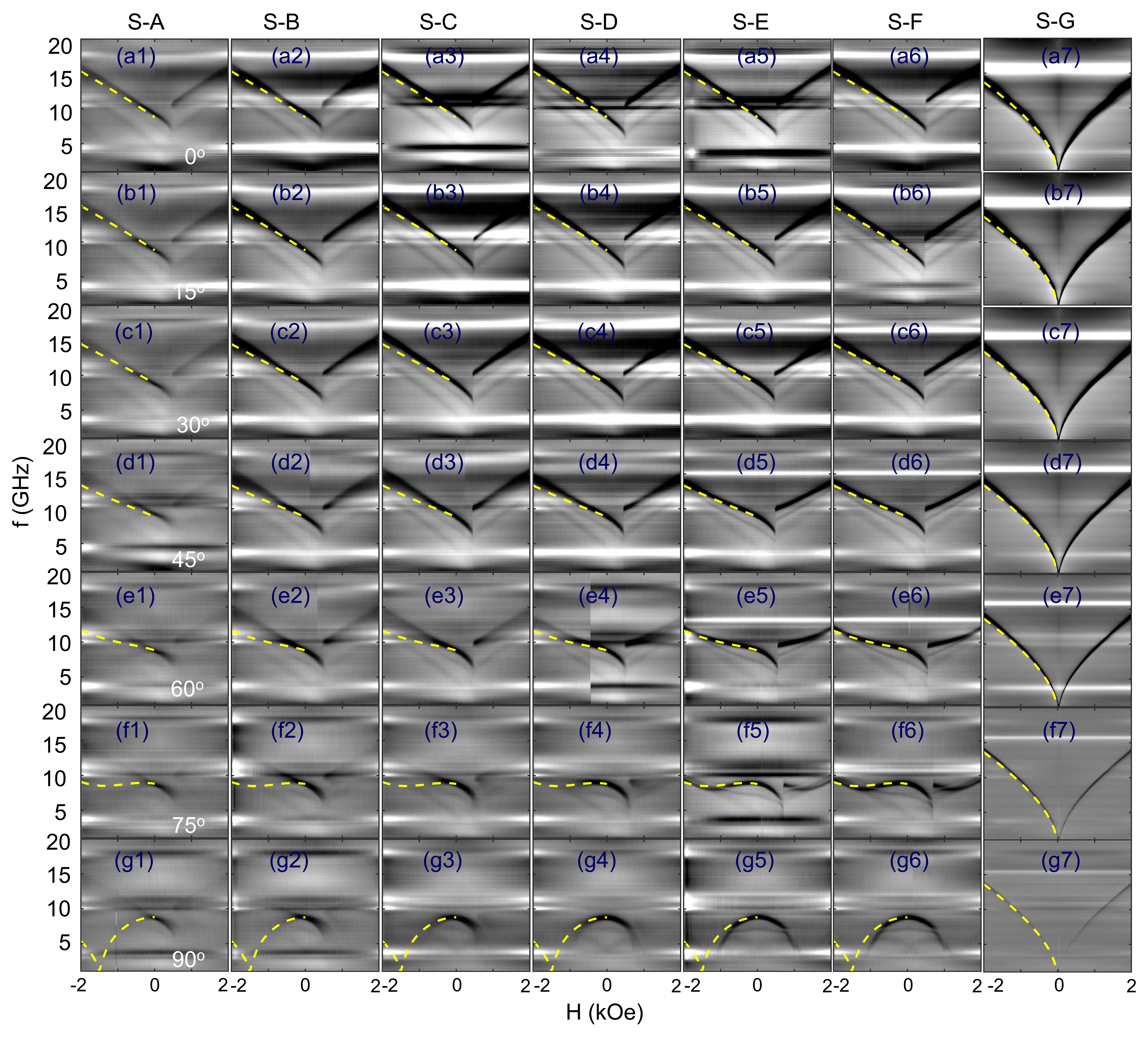}
		\caption{Grayscale VNA-FMR spin-wave spectra  showing high-contrast lines that represent spin-wave branches for samples at three different in-plane field angles. Sub-figures follow the \textquotedblleft LN\textquotedblright naming format, where \textquotedblleft L\textquotedblright represents letters and \textquotedblleft N\textquotedblright represents numbers. Sub-figures with the same letter correspond to the same magnetic field angle (indicated on the right side of the first column), while those with the same number correspond to the same sample (indicated at the top of the first row).   The dashed yellow lines indicate frequency values calculated using Eq. (S3) with $M_{eff} = 1$ T (equivalent to $M_s = 8 \times 10^5$ A/m).  
		}\label{SI1}
	\end{figure} 
	\begin{figure}[htp]
		\includegraphics [width=0.98\textwidth]{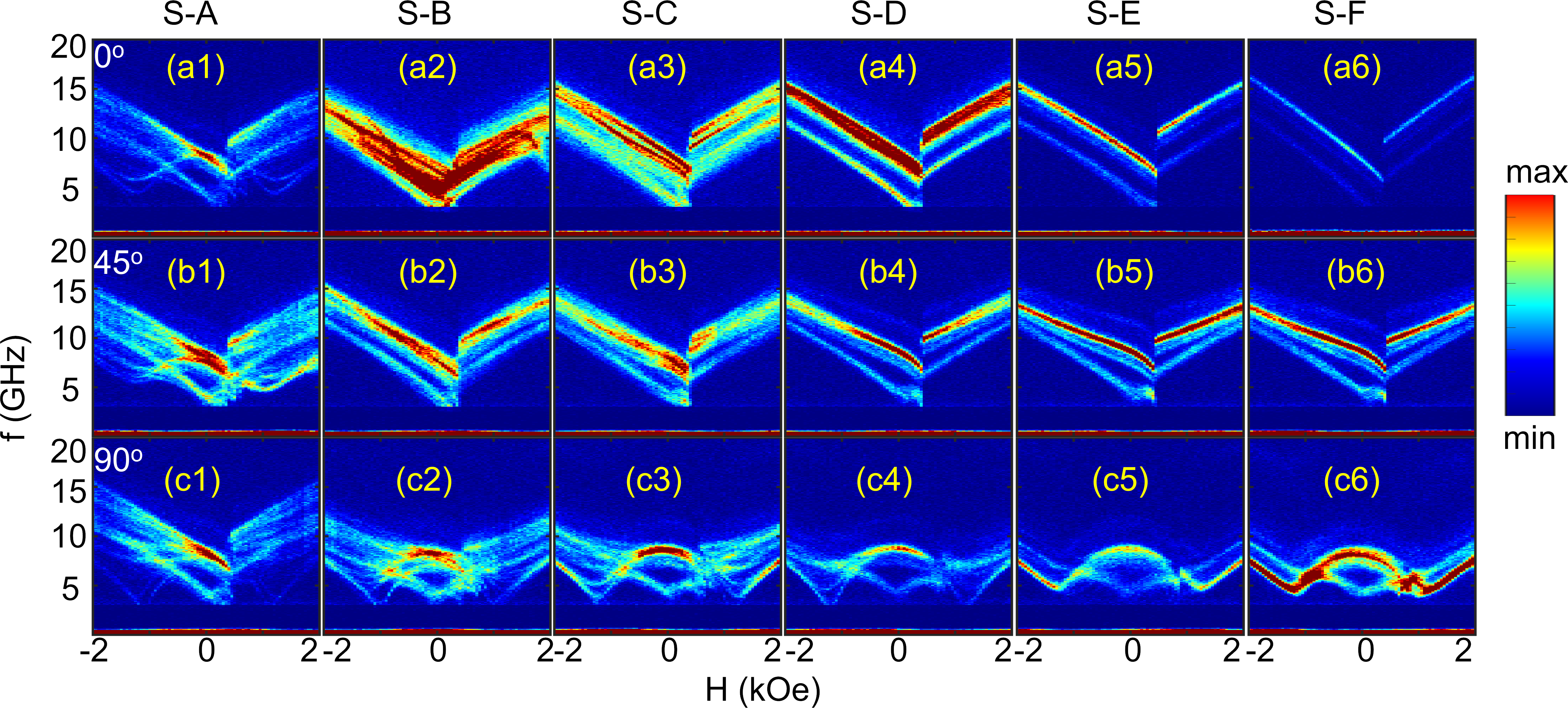}
		\caption{Jet color-scale thermal BLS spectra, showing high-contrast lines that represent spin-wave branches for samples at three different in-plane field angles. Sub-figures follow the \textquotedblleft LN\textquotedblright naming format, where \textquotedblleft L\textquotedblright represents letters and \textquotedblleft N\textquotedblright represents numbers. Sub-figures with the same letter correspond to the same magnetic field angle (indicated on the left side of the first column), while those with the same number correspond to the same sample (indicated at the top of the first row). 
		}\label{SI1_} 
	\end{figure} 
	\begin{figure*}[htp]
		\includegraphics[width=0.98\textwidth]{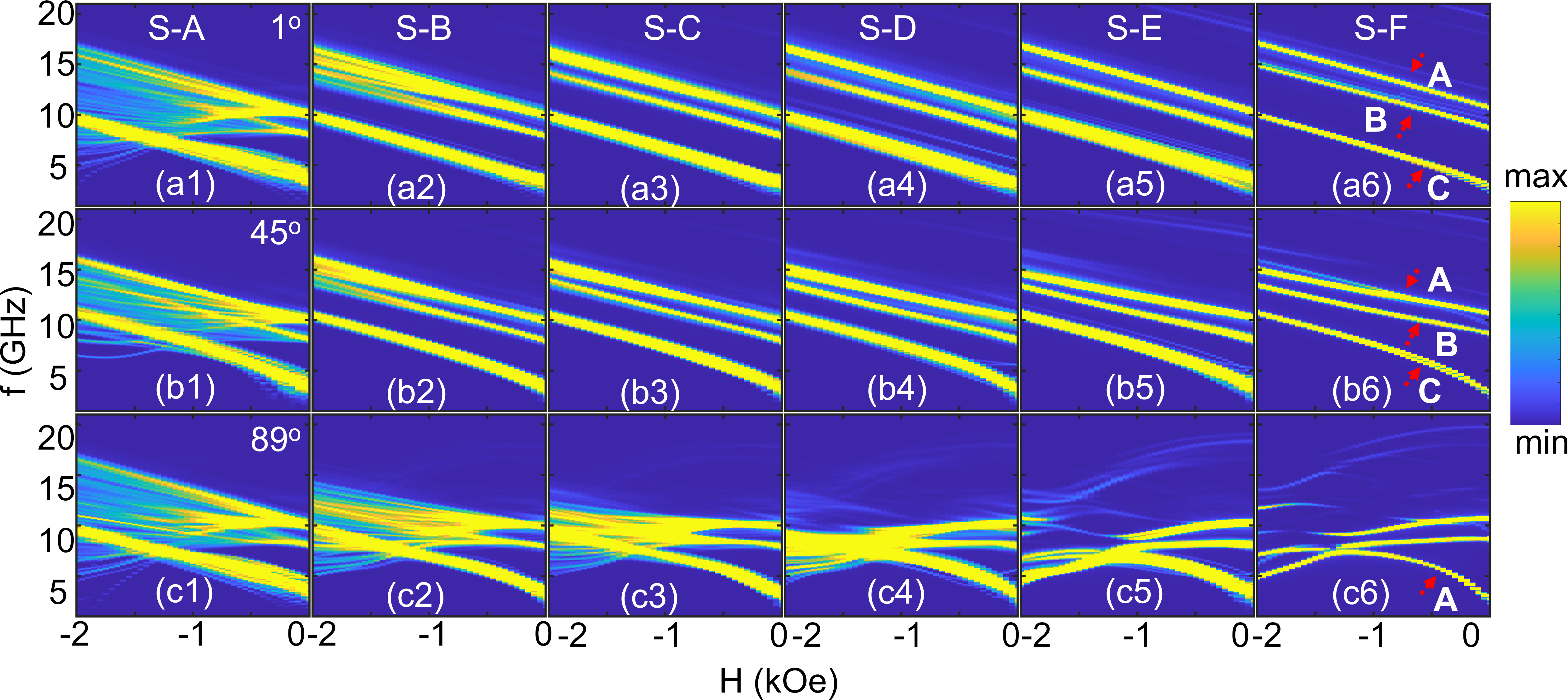}
		\caption{ Integrated micromagnetic simulation power spectra for samples. The color scale indicates square of spin precession amplitudes, with blue representing the minimum and yellow representing the maximum.  Sub-figures follow the \textquotedblleft LN\textquotedblright naming format, where \textquotedblleft L\textquotedblright represents letters and \textquotedblleft N\textquotedblright represents numbers. Sub-figures with the same letter correspond to the same magnetic field angle (indicated on the right side of the first column), while those with the same number correspond to the same sample (indicated at the top of the first row). 
			The arrows and labels in panels (a6), (b6), and (c6) indicate the spin-wave branch designations for sample \textit{S-F} at $\phi_{h} = 1^\circ$, $45^\circ$, and $89^\circ$, respectively.}\label{Fig3_SM} 
		
	\end{figure*} 
	\begin{figure}[htp]
		\includegraphics [width=0.6\textwidth]{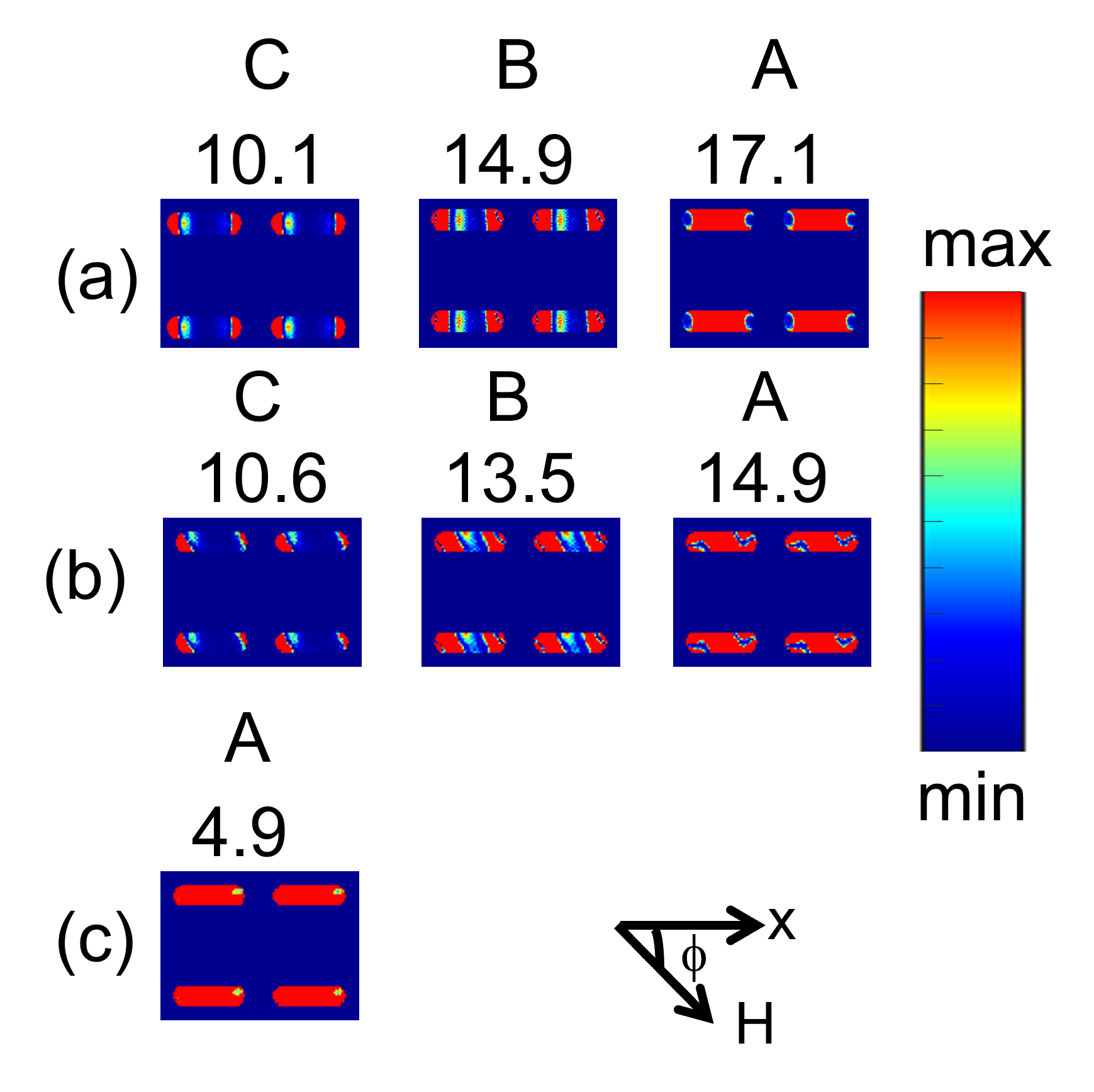}
		\caption{ Local absorption power maps for sample \textit{S-F} at \textit{H} = $-2$~kOe are shown for (a) $\phi_h = 1^\circ$, (b) $45^\circ$, and (c) $89^\circ$. The digits above each map indicate the corresponding resonance frequencies in GHz, while the letters denote the spin-wave branches identified in Fig.~2 of the main text and Fig.~\ref{Fig3_SM}. The color scale ranges from blue (minimum spin precession) to red (maximum spin precession).
		}\label{SI2} 
	\end{figure} 
	
	\clearpage
	
	\subsection{Calculation of Disorder and Complexity Metrics}
	The metrics used to quantify structural disorder and dynamic complexity were calculated based on the nanobar geometries and micromagnetic simulation results as follows. This section provides detailed explanations of the Shannon entropy-based metrics ($S_{\text{config}}$, $S_{\text{connect}}${, and $S_{\text{spectral}}$) and the average local dipolar field ($\langle |B_{\text{local}}| \rangle$)}.
	
	\subsubsection{Introduction to Shannon Entropy for Disorder Quantification}
	Shannon entropy, originally formulated in information theory by C. E. Shannon \cite{Shannon1948Bell}, provides a robust mathematical framework for measuring the uncertainty, randomness, or diversity associated with a probability distribution. In the context of physical systems such as our nanomagnet arrays, this concept can be effectively adapted to quantify various forms of structural and dynamic disorder. If we can define a relevant set of states or configurations for the components of a system (e.g., nanobar orientations, local connectivity environments) and determine the probability distribution over these states, the Shannon entropy $S$ is calculated as:
	\begin{equation}
		S = - \sum_{k=1}^{M} p_k \ln(p_k)
		\label{SM_Shanon} 
	\end{equation}
	where $M$ is the number of possible discrete states (or bins, if a continuous variable is discretized into $M$ intervals), and $p_k$ is the probability of the system (or a component) being in state $k$ (or a variable falling into bin $k$). The logarithm is the natural logarithm (base $e$), and the entropy is thus expressed in units of ``nats''. A higher entropy value signifies greater disorder, increased uncertainty about the state of a randomly chosen component, or a wider variety of states being significantly populated within the ensemble. We apply this principle to calculate $S_{\text{config}}$, $S_{\text{connect}}${, and $S_{\text{spectral}}$}.

	\subsubsection*{Configurational Shannon Entropy ($S_{\text{config}}$)}
	The Configurational Shannon Entropy ($S_{\text{config}}$) quantifies the degree of randomness in the specified orientations of the nanobar long axes. 
	\newline
	\paragraph*{Step-by-Step Calculation for $S_{\text{config}}$}
	\subparagraph*{Step 1: Determine Nanobar Orientations.}
	 For each of the $N$ (\textgreater 60,000) nanobars, {in the arrays used for lithography}, we first determine its orientation vector $\vec{u}_i$ [e.g., from its start coordinate $(x_{i,1}, y_{i,1})$ to its end coordinate $(x_{i,2}, y_{i,2})$]. From this vector $\vec{u}_i = (u_{ix}, u_{iy})$, we calculate the orientation angle $\theta_i$ relative to a reference direction (e.g., the positive $x$-axis). This angle is defined over the full range $[0, 2\pi)$ to capture unique vector directions.
	
	\subparagraph*{Step 2: Create a Probability Distribution of Orientations.}
	We use the set of $N$ orientation angles $\{\theta_i\}$ to form a probability distribution. The range of angles $[0, 2\pi)$ is divided into $M$ equally spaced bins (for our calculations, $M=100$). We then construct a histogram by counting the number of nanobar orientation angles $n_k$ that fall into each bin $k$. The probability $p_k$ for an angle to fall into the $k$-th bin is calculated as:
	\begin{equation}
		p_k = \frac{n_k}{N}
		\label{eq:sm_prob_config_v4}
	\end{equation}
	This $p_k$ represents the fraction of nanobars having an orientation within the angular range defined by bin $k$.
	
	\subparagraph*{Step 3: Calculate Configurational Entropy.}
	Using these probabilities $p_k$, we compute the Configurational Shannon Entropy ($S_{\text{config}}$) using the standard Shannon entropy formula [Eq.~(\ref{SM_Shanon})]:
	\begin{equation}
		S_{\text{config}} = - \sum_{k=1}^{M} p_k \ln(p_k)
		\label{eq:sm_S_config_v4}
	\end{equation}
	The summation is performed over all bins $k$ for which $p_k > 0$. For samples \textit{S-E} and \textit{S-F}, where all nanobars are nominally aligned, $S_{\text{config}}$ is zero.

	\subsubsection*{Connectivity Shannon Entropy ($S_{\text{connect}}$)}
	The Connectivity Shannon Entropy, $S_{\text{connect}}$, quantifies the heterogeneity of the local interaction environments experienced by nanobars within the array. A higher $S_{\text{connect}}$ value indicates a greater diversity of these environments. It is calculated through the following steps:

\subparagraph*{Step 1: Determine Nanobar Properties.} For each of the $N$ nanobars in the array (indexed by $i$), we determine its center coordinates $(x_{c,i}, y_{c,i})$ and the orientation angle of its long axis, $\theta_{\text{axis},i}$. This angle is defined within the range $[0, \pi)$ to represent the undirected nature of the nanobar's axis.

\subparagraph*{Step 2: Define Pairwise Connection Weights.} For every distinct pair of nanobars, $(i,j)$:
(a) Calculate the center-to-center Euclidean distance, $r_{ij}$.
(b) Determine the acute angle difference, $\delta\theta_{ij} \in [0, \pi/2]$, between their respective orientation axes ($\theta_{\text{axis},i}$ and $\theta_{\text{axis},j}$).
(c) Define an effective interaction radius, $R_{\text{eff}}(i,j)$, which depends on the relative alignment of the pair. The radius is set to $R_{\text{eff}} = 550$ nm if the nanobars are closely aligned (i.e., $\delta\theta_{ij} < \pi/20$, or $9^{\circ}$), and $R_{\text{eff}} = 500$ nm otherwise. This reflects a stronger effective interaction range for aligned elements.
(d) Assign a connection weight, $w_{ij}$. If the nanobars are within the effective interaction radius [$0 < r_{ij} \le R_{\text{eff}}(i,j)$], the weight is $w_{ij} = 1/r_{ij}^3$, reflecting a dipolar-like decay of interaction strength. Otherwise, if they are further apart or $r_{ij}=0$, $w_{ij}=0$.

\subparagraph*{Step 3: Calculate Weighted Degree for Each Nanobar.} For each nanobar $i$, compute its weighted degree, $d_{w,i}$. This value represents the cumulative interaction strength experienced by nanobar $i$ from all other nanobars in the array. It is calculated by summing the pairwise connection weights $w_{ij}$ involving nanobar $i$: $d_{w,i} = \sum_{j \neq i} w_{ij}$.

\subparagraph*{Step 4: Create a Probability Distribution of Weighted Degrees.} From the set of $N$ calculated weighted degrees, $\{d_{w,i}\}$, we construct a probability distribution. First, the observed range of $d_{w,i}$ values is discretized into $M'=100$ equally-sized bins. Next, for each bin $k$, we count the number of nanobars, $n'_k$, whose weighted degrees $d_{w,i}$ fall within that bin. The probability $p'_k$ of a randomly selected nanobar having a weighted degree in bin $k$ is then $p'_k = n'_k / N$.

\subparagraph*{Step 5: Calculate Connectivity Entropy.} Finally, using the probability distribution $\{p'_k\}$ derived in Step 4, we compute the Connectivity Shannon Entropy, $S_{\text{connect}}$. This is achieved by applying the standard Shannon entropy formula [as generally defined in Eq.~(\ref{SM_Shanon})]:
\begin{equation}
	S_{\text{connect}} = - \sum_{k=1}^{M'} p'_k \ln(p'_k)
	\label{eq:sm_S_connect_v4}
\end{equation}
The summation is performed over all bins $k$ for which $p'_k > 0$, since $\ln(0)$ is undefined. A higher resulting $S_{\text{connect}}$ value signifies a more heterogeneous distribution of local interaction strengths within the nanobar array.
	\subsubsection*{Average Local Dipolar Field ($\langle |B_{\text{local}}| \rangle$)}		
	This metric estimates the average magnitude of the static dipolar field experienced by a nanobar due to its neighbors, calculated using a point-dipole approximation. Each nanobar $j$ was treated as a point dipole at its center $(x_{c,j}, y_{c,j})$ with magnetic moment $\vec{m}_j$ (magnitude $m = M_s V \approx 3.33 \times 10^{-16} \, \text{A}\cdot\text{m}^2$, aligned with its axis $\theta_j$). The dipolar field $\vec{B}_{ij}$ from $j$ at $i$ is $\vec{B}_{ij} = \frac{\mu_0}{4\pi} ( \frac{3(\vec{m}_j \cdot \hat{r}_{ij})\hat{r}_{ij} - \vec{m}_j}{|\vec{r}_{ij}|^3} )$ \cite{coey2010magnetism}. Contributions from neighbors $j$ where $100 \, \text{nm} \le r_{ij} \le 470 \, \text{nm}$ were summed: $\vec{B}_{\text{local}, i} = \sum_{j \neq i, \text{valid}} \vec{B}_{ij}$. The reported metric is $\langle |B_{\text{local}}| \rangle = \frac{1}{N} \sum_{i=1}^{N} |\vec{B}_{\text{local}, i}|$.
	
	\subsubsection*{Spectral Entropy ($S_{\text{spectral}}$)}	
The spectral data  obtained for each sample (via micromagnetic simulation) was analyzed to identify characteristic features, specifically local maxima (peaks) and inflection points. Prior to feature detection, the power spectrum within the 1 to 20 GHz range was normalized by its maximum value to facilitate consistent parameter application across samples with varying overall intensities. Local maxima were identified directly from this normalized power spectrum using a peak finding algorithm based on criteria such as minimum peak prominence (e.g., $1 \times 10^{-4}$ relative units) and minimum peak height (e.g., $1 \times 10^{-5}$ relative units). {Here, minimum peak height acts as an absolute floor, meaning a peak's value must, at a minimum, reach this specific level (e.g., $1 \times 10^{-5}$ in our case). Minimum peak prominence, conversely, is a measure of a peak's distinctness relative to its immediate surroundings; it quantifies how much the peak rises above the adjacent signal levels before one encounters any higher peak (e.g., by at least $1 \times 10^{-4}$ in our methodology). In our approach, the selected minimum peak prominence ($1 \times 10^{-4}$) is greater than the minimum peak height ($1 \times 10^{-5}$). This configuration ensures that identified peaks are not only above a certain absolute signal level but also represent distinct features that rise significantly above their local baseline, thereby filtering for more structurally significant peaks.} 

Inflection point determination was implemented to capture changes in the spectral curvature, often corresponding to the shoulders of peaks or transitions in the slope. These were identified by first smoothing the normalized power spectrum using a Savitzky-Golay filter (e.g., polynomial order 3, frame length 11 points, with specific adjustments for sample \textit{S-C} to order 4) \cite{savitzky1964smoothing}. The first and second derivatives of this smoothed, normalized spectrum were then computed. Inflection points were located where the second derivative changed sign, subject to filtering criteria: the normalized power at the inflection point had to exceed a minimum threshold (e.g., 0.05 for \textit{S-A,B,D,E,F} and 0.02 for \textit{S-C}), and the absolute value of the normalized first derivative (slope) at that point also had to meet a minimum threshold (e.g., 0.02). This dual-criteria approach for inflection points helps to isolate more significant changes in curvature from minor undulations.
The inclusion of inflection point analysis, alongside traditional peak detection, provides a more comprehensive characterization of the key spectral features {[$M_{N}$ in Fig. 2(c) of the main text]}, highlighting not only the dominant resonant frequencies but also the transitional regions and subtle structural details within the spectra.    
    
    The spectral entropy ($S_{\text{spectral}}$) metric quantifies the complexity of the power distribution among {key spectral features (that is, peaks or inflection points)}  at $H = -200$ mT. Let $P_l$ be the power of the $l$-th {key spectral feature.} These are normalized to $p_l = P_l / \sum P_j$. Then $S_{\text{spectral}} = - \sum p_l \ln(p_l)$ [using Eq.~(\ref{SM_Shanon})]. A higher $S_{\text{spectral}}$ indicates power distributed more evenly among more {key spectral features.}\\

\subsection*{Spatial and Orientational Correlation Functions}
To quantitatively describe the structure of our nanomagnet arrays beyond simple averages, we calculated the pair correlation function, $g_{pc}(r)$, and the orientational correlation function, $g_{oc}(r)$. These functions reveal how the position or orientation of one nanobar influences that of others as a function of their separation distance $r$. Such structural correlations are crucial as they directly impact inter-nanobar interactions and the collective dynamic behavior of the system.
\subsubsection*{Pair Correlation Function, \texorpdfstring{$g_{pc}(r)$}{g\_pc(r)}}
The pair correlation function (PCF), $g_{pc}(r)$, measures the likelihood of finding a nanobar center at a distance $r$ from a reference nanobar center, relative to that expected for a spatially random distribution. It is formally defined through the two-nanobar density $\rho^{(2)}(r)$ for a homogeneous and isotropic system with average number density $\rho = N/A$ (where $N$ is the number of nanobars in a 2D area $A$) as \cite{HansenMcDonald, McQuarrieStatMech}:
\begin{equation} \label{eq:sm_gpc_def_formal}
	\rho^{(2)}(r) = \rho^2 g_{pc}(r)
\end{equation}
Equivalently, $g_{pc}(r)$ can be defined as the ratio of the actual number of distinct pairs of nanobars separated by a distance $r$ to the number of pairs expected at that separation if the system were completely random.

\paragraph*{Analytical Formulation and Computational Implementation.}
Consider an infinitesimal annular shell of radius $r$ and thickness $dr$, with area $A_{\text{shell,infinitesimal}} = 2\pi r dr$. In a 2D system of $N$ nanobars in a total area $A_{\text{total}}$, the expected number of distinct pairs, $dN_{\text{ideal}}(r)$, whose separation distance falls between $r$ and $r+dr$ in a random (ideal gas) configuration is given by:
\begin{equation}
	dN_{\text{ideal}}(r) = \frac{N(N-1)}{2} \frac{A_{\text{shell,infinitesimal}}}{A_{\text{total}}} \approx \frac{N^2}{2A_{\text{total}}} (2\pi r dr) = N \rho \pi r dr
	\label{eq:sm_gpc_analytical_ideal_pairs}
\end{equation}
for large $N$, where $\rho = N/A_{\text{total}}$ is the average number density.
If $dN_{\text{obs}}(r)$ is the actual observed number of distinct pairs of nanobars with separation between $r$ and $r+dr$, then the pair correlation function is:
\begin{equation}
	g_{pc}(r) = \frac{dN_{\text{obs}}(r)}{dN_{\text{ideal}}(r)} = \frac{dN_{\text{obs}}(r)}{N \rho \pi r dr}
	\label{eq:sm_gpc_analytical_from_pairs}
\end{equation}

For practical computation from discrete nanobar coordinates, this is implemented by using finite bin widths:
\begin{enumerate}
	\item \textbf{System Parameters:} In this work, the total area $A_{\text{total}}$ occupied by the nanobar centers is estimated using a \textbf{bounding box}. This bounding box is defined as the smallest axis-aligned rectangle that encompasses all the nanobar centers.	Its area is calculated from the minimum and maximum $x$ and $y$ coordinates of these centers as follows:
	$$ A_{\text{total}} = (x_{\text{max}} - x_{\text{min}}) \times (y_{\text{max}} - y_{\text{min}}) $$
	where $x_{\text{min}} = \min(\{x_c\})$, $x_{\text{max}} = \max(\{x_c\})$, $y_{\text{min}} = \min(\{y_c\})$, and $y_{\text{max}} = \max(\{y_c\})$ are the minimum and maximum $x$ and $y$ coordinates, respectively, found among all nanobar centers $(x_c, y_c)$.
	This estimated area $A_{\text{total}}$ is then used, along with the total number of nanobars $N$, to determine the number density $\rho = N/A_{\text{total}}$.
	\item \textbf{Pair Distance Calculation and Binning:} Distances $r_{ij}$ between all unique pairs of nanobar centers $(i,j)$ are computed. These distances are then binned into discrete annular shells (bins) of finite width $\Delta r$ (e.g., $\Delta r = 20$~nm, up to $r_{\text{max}} = 1500$~nm). Let $N_{\text{obs}}(r_{\text{bin}})$ be the observed count of distinct pairs whose separation $r_{ij}$ falls within a specific bin $k$ (e.g., with inner radius $r_{k,in}$ and outer radius $r_{k,out}$).
	\item \textbf{Shell Area Calculation:} The exact area of each finite annular shell is $A_{\text{shell}}(r_{\text{bin}}) = \pi (r_{k,out}^2 - r_{k,in}^2)$.
	\item \textbf{Expected Random Pairs in Bin:} The expected number of distinct pairs, $N_{\text{ideal}}(r_{\text{bin}})$, in this finite shell for a random distribution is calculated as $\frac{N \rho}{2} A_{\text{shell}}(r_{\text{bin}})$. This uses the large $N$ approximation for the number of pairs ($N^2/2$ rather than $N(N-1)/2$) implicitly through the use of $\rho = N/A_{total}$ in the $\frac{N\rho}{2}$ factor, consistent with the derivation of Eq. (\ref{eq:sm_gpc_analytical_ideal_pairs}) when integrated over a finite shell.
	\item \textbf{Computing $g_{pc}(r)$:} The pair correlation function for the bin is the ratio:
\end{enumerate}
\begin{equation}
	g_{pc}(r_{\text{bin}}) = \frac{N_{\text{obs}}(r_{\text{bin}})}{N_{\text{ideal}}(r_{\text{bin}})} = \frac{N_{\text{obs}}(r_{\text{bin}})}{\frac{N \rho}{2} A_{\text{shell}}(r_{\text{bin}})}
	\label{eq:sm_gpc_computational_final_matched}
\end{equation}

\paragraph*{Interpretation.}
\begin{itemize}
	\item $g_{pc}(r)=0$: Excluded volume around a nanobar.
	\item $0 < g_{pc}(r)<1$: Lower probability of finding a pair of nanobars at distance $r$ than random.
	\item $g_{pc}(r)=1$: Probability same as a random distribution; $g_{pc}(r) \to 1$ as $r \to \infty$ for disordered systems.
	\item $g_{pc}(r)>1$: Higher probability; peaks indicate preferred inter-nanobar distances and structural ordering.
\end{itemize}

\subsubsection*{Orientational Correlation Function, \texorpdfstring{$g_{oc}(r)$}{g\_oc(r)}}
The orientational correlation function, $g_{oc}(r)$, quantifies how the orientations of the long axes of nanobars are correlated for pairs separated by a distance $r$. For our rod-like nanobars, we use a definition based on the second Legendre Polynomial, $P_2(x) = (3x^2 - 1)/2$, which is standard for characterizing nematic-like (quadrupolar) alignment where the axis orientation $\hat{u}_i$ is equivalent to $-\hat{u}_i$ (i.e., the nanobars are ``headless'') \cite{de1993physics, chaikin1995principles, bray1984theory}.

\paragraph*{Rationale and Definition.}
The choice of $P_2(x)$ is due to its suitability for quadrupolar symmetry and its sensitivity to nematic alignment. $P_2[\cos(\delta\theta_{ij})]$ (where $\delta\theta_{ij}$ is the angle between axes $i$ and $j$) is 1 for parallel/anti-parallel alignment and -0.5 for perpendicular alignment. This function effectively probes the $l=2$ (quadrupolar) component of the orientational structure. $g_{oc}(r)$ is the average of $P_2(\cos(\delta\theta_{ij}))$ over all pairs $(i,j)$ separated by distance $r$:
\begin{equation}
	g_{oc}(r) = \left\langle P_2(\cos(\delta\theta_{ij})) \right\rangle_r = \left\langle \frac{3 \cos^2(\delta\theta_{ij}) - 1}{2} \right\rangle_r
	\label{eq:sm_goc_def_final}
\end{equation}
where $\delta\theta_{ij}$ is the acute angle ($[0, \pi/2]$) between the orientation axes of nanobars $i$ and $j$. Nanobar axis orientations are defined by angles $\theta_{\text{axis}} \in [0, \pi)$ relative to a reference direction.

\paragraph*{Computation.}
\begin{enumerate}
	\item Nanobar pairs are binned by their center-to-center distance $r_{ij}$.
	\item For each pair $(i,j)$ in a given distance bin, the acute angle $\delta\theta_{ij}$ between their orientation axes (derived from their individual orientation angles $\theta_i, \theta_j \in [0, \pi)$) is calculated as $\delta\theta_{ij} = \operatorname{acos}(|\hat{u}_i \cdot \hat{u}_j|)$, where $\hat{u}_{ij}$ are unit vectors along the nanobar axes.
	\item $P_2[\cos(\delta\theta_{ij})]$ is computed for each pair.
	\item $g_{oc}(r_{bin})$ for the bin is the average of these $P_2$ values.
\end{enumerate}

\paragraph*{Interpretation.}
\begin{itemize}
	\item $g_{oc}(r)=1$: Perfect parallel or anti-parallel alignment of axes.
	\item $g_{oc}(r)=-0.5$: Perfect perpendicular alignment.
	\item $g_{oc}(r)=0$: Random (uncorrelated) axial orientations.
\end{itemize}
 

	As discussed in the main text [Figs.~2(a) and (b)], $g_{pc}(r)$ shows sharp peaks indicative of long-range order for the periodic sample \textit{S-F}, transitioning to damped oscillations characteristic of short-range order for the positionally disordered samples (\textit{S-E} to \textit{S-A}).
For the orientational correlation, $g_{oc}(r)$ remains close to 1 for samples \textit{S-F} and \textit{S-E} (confirming long-range orientational order), decaying towards lower values for samples \textit{S-D} [to sample \textit{S-A}] as rotational disorder increases, indicating loss of orientational correlation.	
	\subsection*{Correlation and Regression Analysis}
	To further investigate the relationships between the structural disorder metrics and the resulting dynamic complexity, we performed correlation and multiple linear regression analyses on the data obtained for samples S-A through S-F.
	
	\subsubsection*{Correlation Analysis}
\begin{figure}[h!] 
		\centering
		\includegraphics[width=0.6\textwidth]{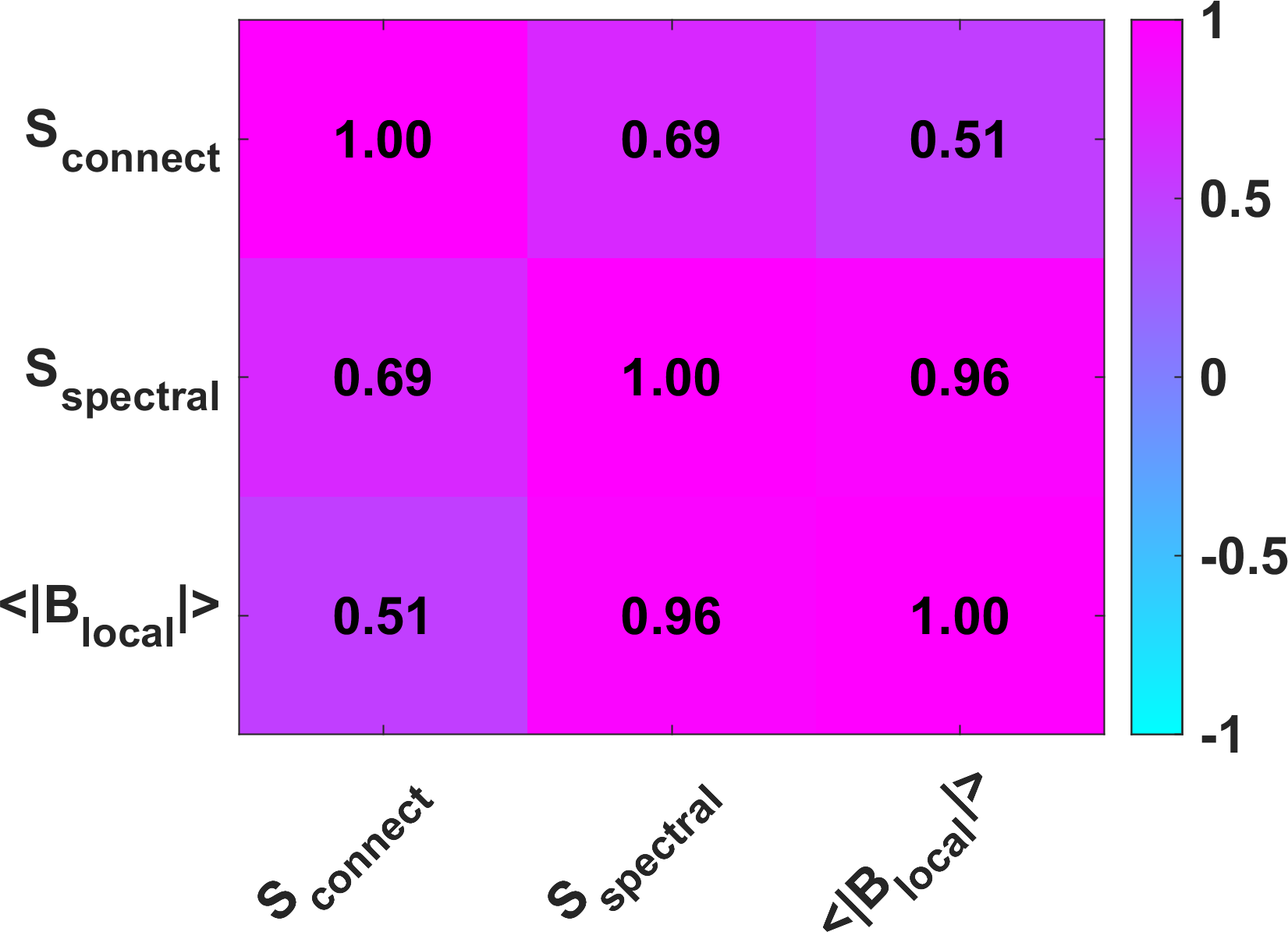} 
		\caption{Pearson correlation matrix heatmap illustrating the linear relationships between connectivity Shannon entropy ($S_{\text{connect}}$), spectral entropy ($S_{\text{spectral}}$), and average local dipolar field ($\langle |B_{\text{local}}| \rangle$) across samples S-A to S-F. Color intensity (ranging from blue/cyan for lower correlation values towards magenta for higher positive correlation values, as per the color bar) and numerical values within each cell indicate the strength of the correlation coefficient.}
		\label{SI5} 
	\end{figure}

	We calculated the Pearson correlation coefficients \cite{SnedecorCochranStatMethods} between the Connectivity Shannon entropy ($S_{\text{connect}}$), the spectral Shannon entropy ($S_{\text{spectral}}$), and the average local dipolar field strength ($\langle |B_{\text{local}}| \rangle$) across all six samples. The resulting correlation matrix is visualized as a heatmap in Fig.~\ref{SI5}. The analysis reveals positive linear correlations among the variables. There is a moderate positive correlation (Pearson coefficient $\approx 0.51$) between $S_{\text{connect}}$ and $\langle |B_{\text{local}}| \rangle$. A strong positive correlation (coefficient $\approx 0.69$) exists between $S_{\text{connect}}$ and $S_{\text{spectral}}$. A very strong positive correlation (coefficient $\approx 0.96$) is observed between $\langle |B_{\text{local}}| \rangle$ and $S_{\text{spectral}}$, suggesting that both structural heterogeneity and average interaction strength are strong linear indicators of the dynamic complexity in this system.
	
	\subsubsection*{Multiple Linear Regression}
We performed a multiple linear regression \cite{KutnerAppliedLinearModels} to model the spectral Shannon entropy ($S_{\text{spectral}}$) as a function of both Connectivity Shannon entropy ($S_{\text{connect}}$) and average local dipolar field ($\langle |B_{\text{local}}| \rangle$). The model equation is:
$S_{\text{spectral}} = \beta_0 + \beta_1 S_{\text{connect}} + \beta_2 \langle |B_{\text{local}}| \rangle$.
The multiple linear regression model, $S_{\text{spectral}} = \beta_0 + \beta_1 S_{\text{connect}} + \beta_2 \langle |B_{\text{local}}| \rangle$, was chosen because results from correlation analysis (discussed above) showed that both $S_{\text{connect}}$ (representing interaction heterogeneity) and $\langle |B_{\text{local}}| \rangle$ (average interaction strength) exhibit strong individual linear correlations with $S_{\text{spectral}}$. The configurational entropy, $S_{\text{config}}$, which specifically quantifies orientational randomness based on Shannon entropy \cite{Shannon1948Bell}, was not included in this particular combined model. Instead, its impact was analyzed separately (Fig.~\ref{SI6}) to clearly isolate the distinct contribution of orientational randomness to spectral complexity. This approach aids in disentangling the effects of different forms of disorder.\\	
	The overall regression model was statistically significant (F-statistic = 46.92, p-value = 0.0054), indicating that the predictors together reliably explain variance in the spectral entropy. The model achieved a very high coefficient of determination (Adjusted $R^2 \approx 0.948$), suggesting that approximately 95\% of the variance in $S_{\text{spectral}}$ across the samples can be linearly accounted for by $S_{\text{connect}}$ and $\langle |B_{\text{local}}| \rangle$ combined. This analysis supports the conclusion that the structural changes, particularly those influencing the average interaction strength and local environment heterogeneity, are strong drivers of the observed increase in dynamic spectral complexity.

	\subsection*{Impact of Rotational Disorder}
\begin{figure}[h!]
		\centering
		\includegraphics[width=0.6\textwidth]{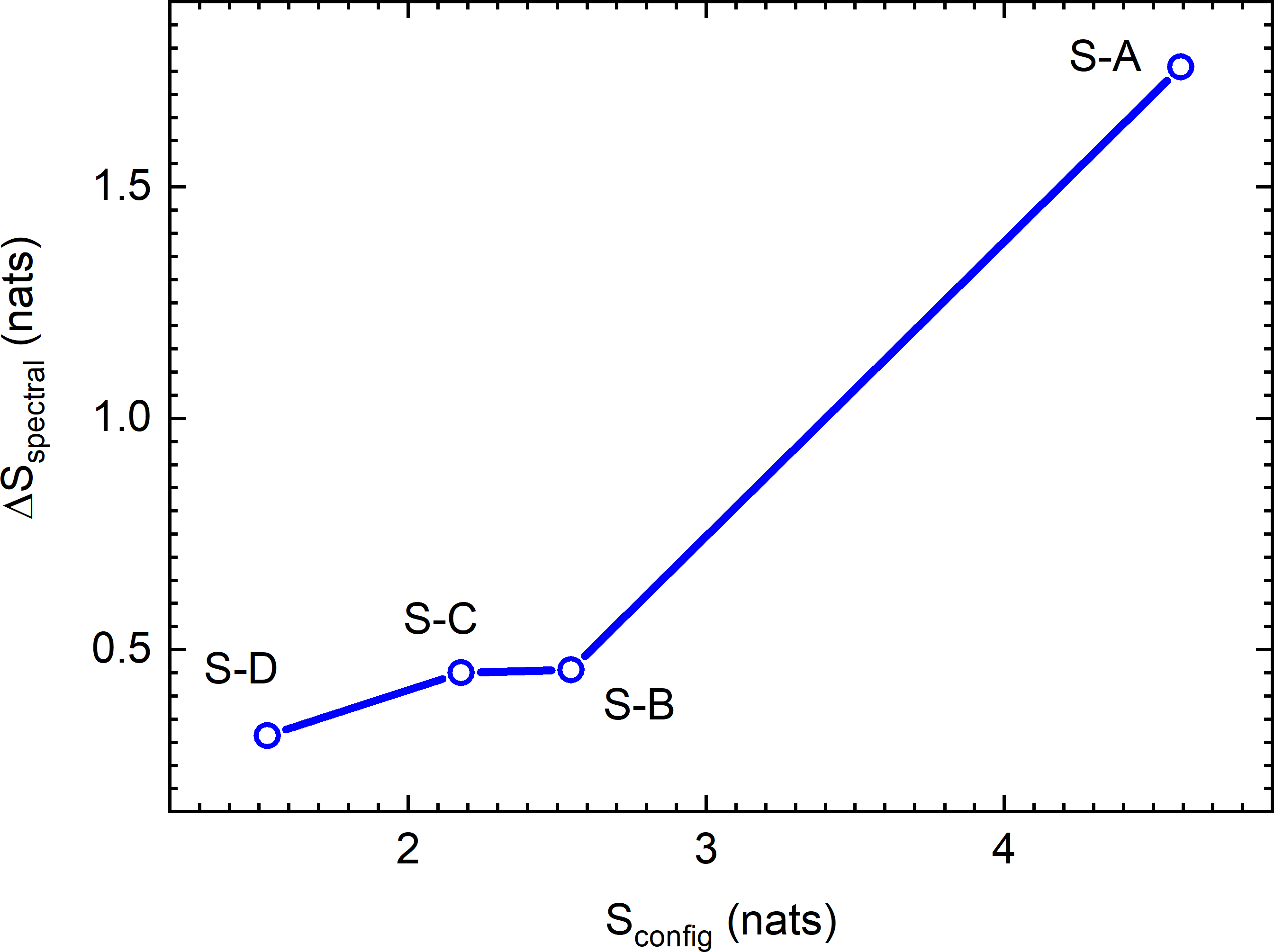} 
		\caption{Impact of rotational disorder on dynamic complexity. The plot shows the change in spectral entropy relative to the positionally disordered sample E ($\Delta S_{\text{spectral}} = S_{\text{spectral}} - S_{\text{spectral, E}}$) as a function of the configurational Shannon entropy ($S_{\text{config}}$) for samples \textit{S-D}, \textit{S-C}, \textit{S-B}, and \textit{S-A}. The clear, approximately linear positive trend highlights how increasing orientational randomness systematically enhances the spectral complexity.}
		\label{SI6} 
	\end{figure}

	To specifically isolate the impact of rotational disorder, we plotted the change in spectral Shannon entropy relative to the positionally disordered sample E ($\Delta S_{\text{spectral}} = S_{\text{spectral}} - S_{\text{spectral, E}}$) against the configurational Shannon entropy ($S_{\text{config}}$) for samples \textit{S-D}, \textit{S-C}, \textit{S-B}, and \textit{S-A} (Fig.~\ref{SI6}). The plot shows a clear, approximately linear positive trend, demonstrating that increasing orientational randomness systematically and significantly enhances the spectral complexity, building upon the baseline complexity introduced by positional disorder.

	\clearpage
	\section{acknowledgment}
    \indent  	
    This material is based upon work supported by the National Science Foundation under Grant No. 2339475. The authors acknowledge the use of facilities and instrumentation supported by NSF through the University of Delaware Materials Research Science and Engineering Center, DMR-2011824. The supercomputing time was provided by DARWIN (Delaware Advanced Research Workforce and Innovation Network), which is supported by NSF Grant No. MRI-1919839. MBJ acknowledges the JSPS Invitational Fellowship for Researcher in Japan.
	\clearpage
	\section{References}
	\bibliographystyle{apsrev4-2} 
	\bibliography{Bibliography1}